\documentstyle[aps,prl,epsf]{revtex}

\begin{document}
\newcommand\beq{\begin{equation}}
\newcommand\eeq{\end{equation}}
\newcommand\bea{\begin{eqnarray}}
\newcommand\eea{\end{eqnarray}}

\def\eps{\epsilon}
\newcommand{\ket}[1]{| #1 \rangle}
\newcommand{\bra}[1]{\langle #1 |}
\newcommand{\braket}[2]{\langle #1 | #2 \rangle}
\newcommand{\proj}[1]{| #1\rangle\!\langle #1 |}
\newcommand{\ba}{\begin{array}}
\newcommand{\ea}{\end{array}}
\newtheorem{theo}{Theorem}
\newtheorem{defi}{Definition}
\newtheorem{lem}{Lemma}
\newtheorem{exam}{Example}
\newtheorem{prop}{Property}
\newtheorem{propo}{Proposition}
\newtheorem{cor}{Corollary}
\newtheorem{conj}{Conjecture}

\twocolumn[\hsize\textwidth\columnwidth\hsize\csname
@twocolumnfalse\endcsname

\author{Peter W. Shor$^*$, John A. Smolin$^\dag$ and Barbara M. Terhal$^\dag$}

\title{Nonadditivity of Bipartite Distillable Entanglement follows from 
Conjecture on Bound Entangled Werner States}

 \address{\vspace*{1.2ex}
            \hspace*{0.5ex}{$^*$AT\&T Research, Florham Park, NJ 07932, USA; 
$^\dag$IBM Watson Research Center,
P.O. Box 218, Yorktown Heights, NY 10598, USA}\\ 
Email: {\tt shor@research.att.com, smolin@watson.ibm.com, terhal@watson.ibm.com}}

\date{\today}

\maketitle
\begin{abstract}
Assuming the validity of a conjecture in Ref. \cite{nptnond1,nptnond2} we 
show that the distillable entanglement for two bipartite states, each of which 
individually has zero distillable entanglement, can be nonzero. We show that
this also implies that the distillable entanglement is not a convex function. Our example consists of the tensor product of a bound entangled state based on an unextendible product basis with an entangled Werner state which lies in the class of conjectured
undistillable states. 
\end{abstract}
\pacs{03.67.Hk, 03.65.Bz, 03.67.-a, 89.70.+c}

]

One of the central goals of the theory of bipartite quantum
entanglement is to develop measures of quantum entanglement. For pure
states, this problem is largely solved. One can formulate a set of
basic requirements \cite{meas:horodecki} which give rise to a unique
measure \cite{vidalem,popror:unique} which is the von Neumann entropy
$S(\rho)=-{\rm Tr} \rho \log \rho$ of the reduced density matrix
$\rho={\rm Tr_A}\,\ket{\psi}\bra{\psi}$ of the pure state
$\ket{\psi}$.  For mixed states, all measures that obey the desirable
requirements have been shown to lie between the regularized
entanglement of formation $E_{\infty}(\rho)=\lim_{n \rightarrow
\infty}E_f(\rho^{\otimes n})/n$ where $E_f(\rho)$ is the
entanglement of formation of $\rho$ \cite{bdsw}, and the distillable
entanglement $D(\rho)$ (see Refs. \cite{bdsw,rainsdist} for proper
definitions of $D$). The special role of $E_{\infty}$ and $D$ among
the possible entanglement measures for mixed states is emphasized by
the fact that they have a direct physical interpretation; they measure
the entanglement costs of making the state $\rho$ asymptotically from
pure states \cite{bdsw,hht:cost} and the amount of pure entanglement
that can be extracted from $\rho$ asymptotically respectively.  Even
though these measures are of central importance in the theory of
bipartite entanglement, various open questions exist about their
basic properties.

There exists 
one class of bipartite density matrices for
which it is known that even though a state $\rho$ in this class is
entangled, the distillable entanglement $D(\rho)=0$. This class of
states is characterized by the fact that the states do not violate the
Peres-Horodecki criterion, i.e.  $({\bf 1} \otimes T)(\rho) \geq 0$,
where $T$ is matrix transposition in a chosen basis. It was shown in
Ref. \cite{pptnodist} that this implies that $D(\rho)=0$. Let us call
these states PPT (``Positive Partial Transpose'') bound entangled states.  Researchers have
considered whether this kind of bound entangled state can play a
role in quantum information processing; for example, it can be proved
that PPT bound entangled states are a useless resource in
protocols of quantum teleportation \cite{teleBE} and also superdense
coding \cite{inprep:horodecki}. 
On the other hand, it has been found that
bound entanglement can be used to quasi-distill a single free
entangled state \cite{BEactivate}, something which is not feasible
without this additional resource. In this Letter we will present an
even stronger effect that bound entangled states can have; states
which are conjectured to be undistillable become distillable by adding
PPT bound entanglement. Let us refer to these conjectured bound
entangled states as NPT (``Negative Partial Transpose'') bound entangled states. This family of
states was considered in Ref. \cite{nptnond1} and \cite{nptnond2}. The
states do violate the Peres-Horodecki criterion, however they seem to
lose this property when trying to squeeze the entanglement (distill) into 
a smaller set of states. Let us state the 
conjecture which was made in Ref. \cite{nptnond1} and \cite{nptnond2}:

\begin{conj}\cite{nptnond1,nptnond2}
Given is the class of Werner states \cite{werner:lhv} in 
${\cal H}_3 \otimes {\cal H}_3$:
\beq
\rho_W(\lambda)=\frac{1}{8 \lambda-1} 
\left(\lambda {\bf 1} - \frac{\lambda+1}{3} H \right).
\eeq
Here $H$ is the swap operator, i.e. $H \ket{i,j}=\ket{j,i}$ for all states 
$\ket{i,j}$ where $i,j=1,\ldots 3$. The state 
$\lim_{ \lambda \rightarrow \infty} \rho_W(\lambda)$ is separable and 
for any finite 
$\lambda \geq 0$ $\rho_W(\lambda)$ is entangled and violates the 
Peres-Horodecki criterion. It is conjectured that 
for all  $\lambda \ge 2$ the state $\rho_W(\lambda)$ is undistillable, 
i.e. $D(\rho_W(\lambda))=0$.
\end{conj}

Before reviewing the evidence for this conjecture, let us recall the
condition for distillability:

\begin{theo}\cite{pptnodist,nptnond1,nptnond2} 
The density matrix $\rho$ is distillable, i.e. $D(\rho) > 0$,  if and only if 
there exists an $n >0$ such that 
\beq
{\rm Tr}\left( \ket{\psi_2}\!\bra{\psi_2} ({\bf 1} \otimes T)(\rho^{\otimes n}) \right) < 0,
\label{neg}
\eeq
where $\ket{\psi_2} \in {\cal H}_A \otimes {\cal H}_B$ is a state with 
Schmidt rank 2 and $T$ is matrix transposition in any basis. 
\label{theo1}
\end{theo}

The evidence in support of the conjecture is the following. If we set
$n=1$ in Theorem 1, one can prove that Eq.~(\ref{neg}) is nonnegative
for all states $\rho_W(\lambda)$ with $\lambda \geq 2$. Furthermore, for $n=2$ 
and $n=3$ numerical evidence for the nonnegativity for Eq.~(\ref{neg})
has been found for these states. 
Also it has been proved that for every finite $n$ in Theorem 1, there
exists a finite $\lambda$
for which Eq. (\ref{neg}) is not satisfied.
The evidence, even though it is convincing, is not conclusive.

In this letter we consider the distillability properties of a pair of 
states, one of which has PPT bound entanglement and one which has 
(conjectured) NPT bound entanglement. Surprisingly we find that 
the distillable entanglement of the pair can be nonzero. Thus 
\beq
D(\rho_1 \otimes \rho_2) > 0,\;\; D(\rho_1)=0, \; D(\rho_2)
\stackrel{\tiny \rm conjectured}{=}0, 
\label{nonaddd}
\eeq
which is an extreme example of {\em nonadditivity} of the distillable entanglement
known as ``superactivation'' \cite{sst:act} assuming that the conjecture holds. 
Examples of superactivation of bound entanglement have previously been found in a 
multipartite system \cite{sst:act}. The strict superaddivity that we find here, 
seems even more surprising since we expect that fewer incomparable resources and 
states exist in the bipartite case.

This nonaddivivity has an added consequence, namely the entanglement measure $D$ 
will not be convex if Conjecture 1 holds. Let us take the states $\rho_1$ 
and $\rho_2$ for which Eq. (\ref{nonaddd}) holds and mix them in this 
way: 
\beq
\rho=\frac{1}{2}\rho_1 \otimes 
(\ket{1}\bra{1})_A+\frac{1}{2}\rho_2 \otimes (\ket{2}\bra{2})_A.
\label{eqconvexity}
\eeq
Convexity of $D$ would imply that 
\bea
D(\rho) \leq \frac{1}{2}D(\rho_1 \otimes (\ket{1}\bra{1})_A)+\frac{1}{2}D(\rho_2 \otimes (\ket{2}\bra{2})_A)  \nonumber \\ \stackrel{\tiny \rm conjectured}{=} 0.
\eea
However, we can show that $D(\rho)>0$.  
To distill the mixture, Alice first 
measures the label $\ket{1}$ and $\ket{2}$ on many copies of $\rho$. This will give 
Alice and Bob a supply of both $\rho_1$ as well as $\rho_2$ which can be distilled since 
$D(\rho_1 \otimes \rho_2) > 0$. We must conclude that demanding 
convexity of an entanglement measure, as was done in Ref. \cite{meas:horodecki}, 
is too constraining \cite{truemixing}.

Another consequence of the result is a nonzero lower bound on the entanglement 
of formation of $\rho_2$. From Proposition 3 in Ref. \cite{upb2} we have 
that the distillable entanglement of $\rho_2$, assisted by bound entanglement 
(e.g. state $\rho_1$ for which $D(\rho_1)=0$) is a lower bound for 
the regularized entanglement of formation $E_{\infty}(\rho_1)$, or 
$E_{\infty}(\rho_1) \geq D(\rho_1 \otimes \rho_2) > 0$. And, note 
that if Conjecture 1 holds, the same is true for state $\rho_1$, i.e. 
$D(\rho_1)=0$, but $E_{\infty}(\rho_1) \geq D(\rho_1 \otimes \rho_2) > 0$ 
which would provide an additional example of irreversible asymptotic entanglement processing \cite{horo:irrev}. 

The distillable state $\rho=\rho_1 \otimes \rho_2$ also provides the first 
nontrivial example of a density matrix which satisfies the reduction criterion
\cite{filterhor}, i.e. ${\bf 1}_A \otimes \rho_B-\rho \geq 0$ and 
$\rho_A \otimes {\bf 1}_B-\rho \geq 0$, while it is distillable. This follows
from the fact that both $\rho_1$ and $\rho_2$ satisfy the reduction criterion
(otherwise they would be distillable) and the fact that any tensorproduct of
states that by themselves satisfy the criterion, satisfies the reduction 
criterion as well \cite{filterhor}. 

For our PPT bound entangled state we choose a bound entangled state
in ${\cal H}_3 \otimes {\cal H}_3$ based on an unextendible product basis
(UPB) \cite{upb1}. In particular, in Ref. \cite{upb1} the ${\bf Pent}$ UPB 
was introduced and the corresponding bound entangled state $\rho_{\bf Pent}$. 
The unextendible product basis is given by five vectors 
\beq
\ket{v_i \otimes v_{2 i\bmod 5}},\;i=0,\ldots,4,
\label{upbs}
\eeq
where
\beq
\ket{v_i}=N (\cos(2 \pi i/5),\sin(2 \pi i/5),h),
\eeq
and $N=2/\sqrt{5+\sqrt{5}}$ and $h=\frac{1}{2}\sqrt{1+\sqrt{5}}$. The bound entangled state 
$\rho_{\bf Pent}$ is equal to
\beq
\rho_{\bf Pent}=\frac{1}{4}({\bf 1}-\sum_{i=0}^4 \ket{v_i,v_{2i \bmod 5}}\bra{v_i,v_{2i \bmod 5}}).
\eeq
Our choice of the NPT bound entangled state is the Werner state $\rho_W(\lambda)$ 
in ${\cal H}_3 \otimes {\cal H}_3$. The partial transpose of this state is
\beq
({\bf 1} \otimes T)(\rho_W(\lambda))=\frac{1}{8 \lambda-1} 
\left(\lambda {\bf 1} -(\lambda+1)\ket{\Psi}\bra{\Psi} \right),
\eeq
where $\ket{\Psi}=\frac{1}{\sqrt{3}}\sum_i \ket{ii}$.

We will show that there exists a vector $\ket{\psi_2}$ which has Schmidt 
rank 2 with the property 
\beq
{\rm Tr}\left( \ket{\psi_2}\!\bra{\psi_2} [{\bf 1} \otimes T](\rho_W(\lambda) \otimes 
\rho_{\bf Pent})\right) < 0,
\label{negexp}
\eeq
for a certain range in $\lambda$. From Theorem \ref{theo1} it then follows 
that $\rho_W(\lambda) \otimes \rho_{\bf Pent}$ is distillable. 
The vector $\ket{\psi_2} \in {\cal H}_{A_1,B_1} \otimes {\cal H}_{A_2,B_2}$ 
can be parametrized as 
\beq
\ket{\psi_2}=\sum_{i,j} \ket{i,j} \otimes \ket{\psi_{ij}},
\eeq
where the vectors $\ket{\psi_{ij}}$ are of the form 
\beq
\ket{\psi_{ij}}=\ket{x_i} \otimes \ket{y_j}+\ket{z_i} \otimes \ket{u_j},
\eeq
due to the fact that $\ket{\psi_2}$ has Schmidt rank 2 over a cut in 
$A_1, A_2$ versus $B_1$ and $B_2$. Here the vectors $\ket{x_i},\ket{y_i},
\ket{z_i},\ket{u_i}$ are unnormalized arbitrary vectors in ${\cal H}_3$, to 
be fixed later. We will not be concerned with the normalization of the vector $\ket{\psi_2}$ 
since this is irrelevant for the sign in Eq. (\ref{negexp}). 

It was noted in Ref. \cite{upb1} that the density matrix $\rho_{\bf Pent}$ is invariant under partial transposition ${\bf 1} \otimes T$. Using this fact and
the parametrization of $\ket{\psi_2}$ we can express Eq. (\ref{negexp}) (dropping the factor $1/(8 \lambda -1)$) in terms of 
the vectors $\ket{\psi_{ij}}$:
\beq
\ba{l}
{\rm Tr}\left[\lambda \sum_i \ket{\psi_{ii}}\bra{\psi_{ii}}-\frac{\lambda +1}{3}\sum_{i,j} \ket{\psi_{ii}}
\bra{\psi_{jj}}\right]\, \rho_{\bf Pent}+\\
\lambda {\rm Tr} \sum_{i \neq j} \ket{\psi_{ij}}\bra{\psi_{ij}}\,\rho_{\bf Pent}.
\ea
\label{negexp2}
\eeq
We make a choice for the vectors $\ket{\psi_{ij}}$ which 
results in 
\beq
\ba{l}
\ket{\psi_2}=2 \ket{0,2} \otimes \ket{v_4,v_3}+
\frac{1}{2} \ket{2,0} \otimes \ket{v_3,v_1}+\\
2 \ket{1,2} \otimes \ket{v_1,v_2}+
\frac{1}{2} \ket{2,1} \otimes \ket{v_2,v_4}+\\
            \ket{0,0} \otimes \ket{v_4,v_1}
            -\ket{1,1} \otimes \ket{v_1,v_4}+ \\
           \ket{2,2} \otimes (\ket{v_3,v_3}-\ket{v_2,v_2}).
\ea
\label{choice2}
\eeq
It can easily be checked that this choice corresponds to setting 
\beq
\left\{\ba{l}
-\ket{x_1}=\ket{z_1}=2 \ket{y_0}=2 \ket{u_0}=\ket{v_1} \\
\ket{x_0}=\ket{z_0}=2 \ket{y_1}=-2 \ket{u_1}=\ket{v_4} \\
2 \ket{x_2}=\ket{u_2}=\ket{v_3}+\ket{v_2} \\
2 \ket{z_2}=\ket{y_2}=\ket{v_3}-\ket{v_2}.
\ea\right.
\eeq

Now we observe the consequences for Eq. (\ref{negexp2}). Since 
we have chosen the states $\ket{\psi_{ij}}$ for $i \neq j$ to be either equal 
to the zero-vector or to one of the UPB vectors, Eq. (\ref{upbs}), we ensure 
that the last term in Eq. (\ref{negexp2}) is 0. We then use that the inner products 
of the remaining vectors with respect to the state $\rho_{\bf Pent}$ are given by 
\bea
\bra{v_1,v_4} \rho_{\bf Pent} \ket{v_1,v_4}=\frac{\sqrt{5}}{2}-1, \nonumber \\
\bra{v_4,v_1} \rho_{\bf Pent} \ket{v_4,v_1}=\frac{\sqrt{5}}{2}-1, \nonumber \\
\bra{v_4,v_1} \rho_{\bf Pent} \ket{v_1,v_4}=\frac{-7+3 \sqrt{5}}{8}, \nonumber \\
\bra{\psi_{22}} \rho_{\bf Pent} \ket{\psi_{22}}=
\sqrt{5}-2-\frac{3-\sqrt{5}}{4}, \nonumber \\
\bra{\psi_{22}} \rho_{\bf Pent} \ket{v_4,v_1}=\frac{-2+\sqrt{5}}{4}, \nonumber \\
\bra{\psi_{22}} \rho_{\bf Pent} \ket{v_1,v_4}=\frac{2-\sqrt{5}}{4}.
\eea
Here $\ket{\psi_{22}}=\ket{v_3,v_3}-\ket{v_2,v_2}$. Hence it follows that Eq. (\ref{negexp2}) equals 
\bea
\bra{\psi_2} [{\bf 1} \otimes T](\rho_W(\lambda) \otimes \rho_{\bf Pent}) \ket{\psi_2}= \nonumber \\
\frac{1}{12}(\lambda (17 \sqrt{5}-37)+20-10 \sqrt{5}).
\eea
This expression is negative when 
\beq
\lambda  < \frac{10 \sqrt{5}-20}{17 \sqrt{5}-37} \approx 2.3300.
\eeq
Thus this solution provides a proof that in the range $\lambda \in [2, 2.3300)$
the state $\rho_{\bf Pent} \otimes \rho_W(\lambda)$ is distillable.

\begin{figure}
\begin{center}
\epsfxsize=8.5cm 
\epsffile{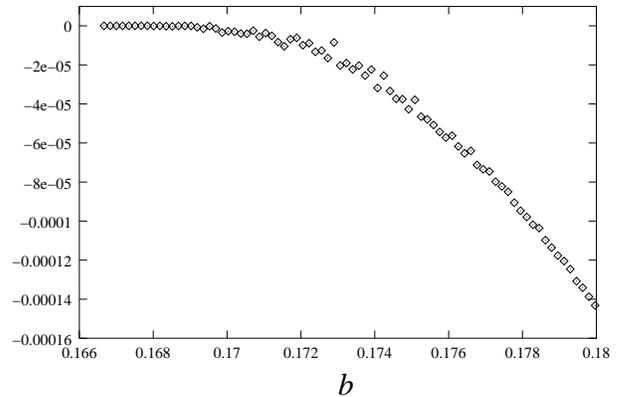}
\caption{Numerical results on the value of Eq. (\ref{negexp}) versus $b$. At $b=1/6$ the 
density matrix $\rho_W(\lambda)$ is separable. When $b > 1/5$ $\rho_W(\lambda)$ is distillable.}
\label{fig1}
\end{center}
\end{figure}

The solution that we have constructed analytically may not be optimal.
We have carried out a numerical study, see Fig. \ref{fig1},  evaluating the 
minimum value of Eq. (\ref{negexp}) 
while varying the parameter $b$ which is related to $\lambda$ by 
$\lambda=(b+1/3)/(8b-4/3)$, or $b \in (1/6,1/5]$ when $\lambda \in (\infty, 2]$. 
As the figure shows, the 
activation effect is extremely small (all density matrices and states are 
normalized here, unlike in the analytical procedure above) and seems to vanish 
before we reach the boundary with the set of separable Werner states (see 
also \cite{extranum}). It is possible that by using two or more states 
$\rho_{\bf Pent}$ for the activation we obtain a negative expectation value 
for smaller values of $b$. 


The activation of $\rho_W(\lambda)$ by $\rho_{\bf Pent}$ is not an
effect particular to $\rho_{\bf Pent}$. The strategy to minimize
Eq. (\ref{negexp2}) can very likely be generalized to other bound
entangled states based on unextendible product bases. We can always
put the last term to zero, by choosing the states $\ket{\psi_{ij}}$ to
be either 0-vectors or UPB-vectors. This gives us some
additional constraints for the states $\ket{\psi_{ii}}$, but the
number of free parameters will be still quite large.

In conclusion, pending the proof of conjecture 1, we have determined
an essential new and surprising property of the distillable
entanglement, namely its capacity to be nonadditive. It is clear that
it would be highly desirable to prove the conjecture, but that goal
remains elusive for the moment.

{\bf Acknowledgments}: The authors would like to thank David DiVincenzo, Pawe\l{} Horodecki and Debbie Leung for interesting discussions. JAS and BMT acknowledge support 
of the ARO under contract No. DAAG-55-98-C-0041.

\bibliographystyle{hunsrt}
\bibliography{refs}

\begin{thebibliography}{10}
\bibitem{extranum}
We have carried out an additional numerical study which indicates there must
  exist a neighborhood at the boundary of the set of separable states (the
  point $b=1/6$), for which Eq. (\ref{negexp}) is nonnegative.
\end{thebibliography}

\end{document}